# An Introduction to Community Detection in Multi-layered Social Network

Piotr Bródka[1], Tomasz Filipowski[1,2], Przemysław Kazienko[1,2]

[1] Wrocław University of Technology, Wyb. Wyspiańskiego 27, 50-370 Wrocław, Poland
[2] Research Engineering Center Sp. z o.o., ul. Strzegomska 46B, 53-611 Wrocław, Poland
`piotr.brodka@pwr.wroc.pl, tomasz.filipowski@pwr.wroc.pl, kazienko@pwr.wroc.pl`

**Abstract:** Social communities extraction and their dynamics are one of the most important problems in today's social network analysis. During last few years, many researchers have proposed their own methods for group discovery in social networks. However, almost none of them have noticed that modern social networks are much more complex than few years ago. Due to vast amount of different data about various user activities available in IT systems, it is possible to distinguish the new class of social networks called multi-layered social network. For that reason, the new approach to community detection in the multi-layered social network, which utilizes multi-layered edge clustering coefficient is proposed in the paper.

## 1   Introduction

Area of science which in recent years is rapidly growing is social network analysis. One of the main reasons for this is growing number of different social networking systems and simple way to obtain data from which we can extract those social networks. Depending on the type of social network we want to build, data can be found in various places, e.g.: blogs [1], telecommunication data [2], bibliographic data [3], social services like Facebook [9], e-mail systems [10] and more.

Group extraction is among those topics which arouse the greatest interest in the domain of social network analysis. Finding groups among the system users opens the new possibilities and can be utilize in such disciplines as human resource management, advertisement, information propagation and lots of others.

Many methods and algorithms have been proposed[3], [8].They are becoming faster [2] and better reflect the real groups in social network [6]. These upgrades have to be significant since social networks are growing very fast those days. However few years ago social networks stopped to grow only in "width", and also began to grow "up". While information technologies became more and more popular, people have expanded their everyday communication on many new channels. They start to build their social environments basing on various web activities.  Social networks have become multi-layered (*MSN*).

*MSN* is a kind of social network where people are connected by many different relationships. Some examples may be complex social networking sites (e.g. Facebook), where people are linked as friends, via common games, "like it", etc. or regular companies where people are at the same time: department colleagues, best friends, colleagues from the company football team, etc. The existing community extraction methods do not concern this multidimensionality of human relations and  allow only to group each layer separately. They do not allow to look at the social network as a whole and extract multi-layered groups based on all network layers. For that reason, the new approach to community detection in multi-layered social networks is proposed in the paper.

## 2   Multi-layered Social Network

**Definition 1**: A multi-layered social network (*MSN)* is defined as a tuple $<V,E,L>$, where: $V$ – is a not-empty set of nodes (social entities: humans, organizations, departments etc.); $E$ – is a set of tuples $<x,y,l>$, $x,y \in V$, $l \in L$, $x \neq y$ and for any two tuples $<x,y,l>$, $<x',y',l'> \in E$ if $x=x'$ and $y=y'$ then $l \neq l'$; $L$ – is a set of distinct layers.

Each tuple $<x,y,l>$ is an edge from $x$ to $y$ in layer $l$ in the multi-layered social network (MSN). The condition $x \neq y$ preserves from loops, i.e. reflexive relations from $x$ to $x$. Moreover, there may exist only one edge $<x,y,l>$ from $x$ to $y$ in a given layer $l$. That means any two nodes $x$ and $y$ may be connected with up to $|L|$ (cardinality of a set $L$) edges coming from different layers. Edges in *MSN* are directed and for that reason, $<x,y,l> \neq <y,x,l>$. Each layer corresponds to one type of relationship between users [4]. Examples of different relationship types can be real world friendship, Facebook friendship, family bonds or work ties. A separate relationship can also be defined based on distinct user activities towards some 'meeting objects', for example publishing photos, commenting photos, adding photos to favourites, etc. See [4], for details. Depending on variety of user activities types, the *MSN* will have more or less layers.

Piotr Bródka, Tomasz Filipowski, Przemysław Kazienko

Nodes $V$ and edges $E$ from only one layer $l \in L$ (such edges form set $E_l$) correspond to a simple, one-layered social network $SN <V, E_l, \{l\}>$.

A multi-layered social network $MSN=<V,E,L>$ may be represented by a directed multi-graph. In Figure 1, the example of three-layered social network is presented. The set of nodes consists of $\{x, y, u, v, z, t\}$ so there are five users in the network that can be connected with each other within three layers: $l_1$, $l_2$ and $l_3$. In layer $l_1$, eight relationships between users: $<x,y,l_1>$, $<y,x,l_1>$, $<x,z,l_1>$, $<z,x,l_1>$, $<y,z,l_1>$, $<u,z,l_1>$, $<u,v,l_1>$, $<v,u,l_1>$ can be distinguished.

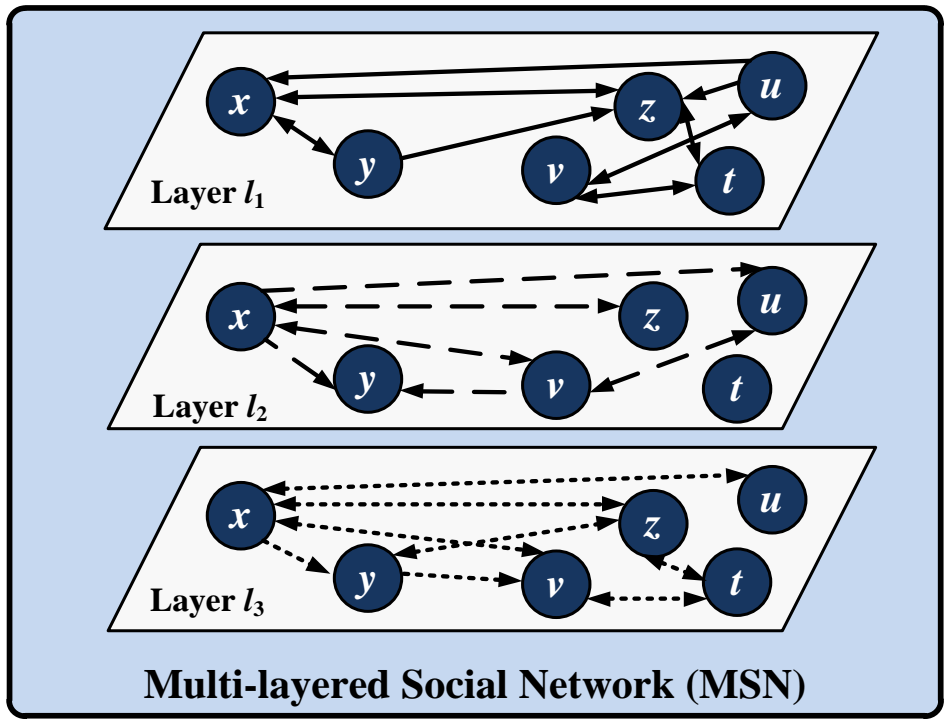

Fig. 1. An example of the multi-layered social network MSN

## 3  Social Community

A group, often also called a social community, in the biological terminology is a number of cooperating organisms, sharing a common environment. In sociology, in turn, it is usually defined as a group of people living and cooperating on a particular area. However, due to the rapid growth of the Internet, telecommunication and transportation, the concept of community has lost its geographical limitations. Overall, a general idea of the social community has become a set of people in a social network, whose members more frequently collaborate with each other rather than with members of this social network who do not belong to the group.

This concept of social community can be easily transposed to the graph theory, in which the social network is represented by a graph. Group is a set of vertices with high density of edges between them, and low edge density between those vertices which do not belong to this set.

However, the problem appears in the quantitative definition of community. There is no general definition of the group (social community) in social networks [8]. There are several of them, which are used depending on the authors' needs.

Most definitions are built based on the idea presented above. Nevertheless, none of them has been commonly accepted. Additionally, groups can also be algorithmically determined, as the output of the specific clustering algorithm, i.e. without a precise a priori definition [5]. In such case, group definition is determined by the algorithm and its parameters. In this paper, we will use such definition.

In this article, we assume that the group $G$ in the social network $SN=<V,E>$ is a subset of vertices from $V$ ($G \subseteq V$), extracted using any community extraction method (clustering algorithm) and the multi-layered group $MG$ in multi-layered social network *MSN=<V,E,L>*. is a subset of vertices from *V* (*MG⊆V*) extracted using community extraction method which takes into account all layers of *MSN*.

## 4   Multi-layered Edge Clustering Coefficient

Before the new measure can be introduced the concept of multi-layered neighbourhood has to be presented. The neighbourhood *N(x)* of a given node *x* for regular one-layered social network SN=<V,E> is defined as:

$$N(x,l) = \{y :< y,x,l >\in E \vee < x,y,l >\in E\} \quad (1)$$

In other words, all nodes connected with the particular node by any edge in the given layer belong to this node neighbourhood in this layer.

Multi-layered neighbourhood, *MN(x,α)*, of a given node *x∈V* is a set of nodes, which are neighbours of node *x* on at least α layers in the *MSN*:

$$MN(x,\alpha) = \{y : |\{l :< x,y,l >\in E \vee < y,x,l >\in E\}| \geq \alpha\} \quad (2)$$

So a node is a muli-layered neighbour of the given node if they are connected by any edges in at least α layers.

For MSN from Figure 1, we have *MN(x,1)={u,v,y,z}*, *MN(x,2)={u,v,y,z}* *MN(x,3)={u,y,z}*

The cross-layer edge clustering coefficient (*CLECC*) is an edge measure which was developed based on idea of edge clustering coefficient measure introduced by Radicchi et. al [7].

Edge clustering coefficient for an edge *<x,y>* expresses how much the neighbours of the user *x*, and neighbours of the user *y* are interconnected. The edge clustering coefficient is defined as:

$$ECC(x,y) = \frac{z_{x,y} + 1}{s_{x,y}} \quad (3)$$

where *x* and *y* are the users connected by the edge *<x,y>*, $z_{x,y}$ is the number of triangles built upon the edge *<x,y>* and all edges between x, y and their neighbours, $s_{x,y}$ is the possible number of triangles that one could build based on edge<x,y> and all possible edges (even those that do not exist) between *x, y* and their neighbours.

*CLECC* measure is slightly modified and express the similar neighbours interconnectivity but for the multi-layered social network.

$$CLECC(x,y,\alpha) = \frac{|MN(x,\alpha) \cap MN(y,\alpha)|}{|(MN(x,\alpha) \cup MN(y,\alpha))/\{x,y\}|} \quad (4)$$

Thus, it can be described as a proportion between the common multi-layered neighbours and all multi-layered neighbours of *x* and *y*

*CLECC,* by utilizing multi-layered neighborhood, considers all layers at the same time. The α parameter allows to adjust the measure strictness depending on differences in the density of each layer. This latter feature is particularly important when there are very large differences in the density of each layer.

For instance, imagine that we have four layers and 1000 users. Two of them are very dense (50,000 edges), while the two other quite sparse (5,000 edges). Now, thanks to the α parameter the measure can be adjust. It can be very restrictive and require the connection to exist on all layers (α=4), or it can be gentle and require connections only in few of them (α=2). We can also choose a middle ground and assume that the connection exists on two existed dense layers and one of the sparse (α=3).

## 5   Community Detection in Multi-layered Social Network

After introducing the *CLECC* measure the method for community detection in multi-layered social network can be presented:

**Input:**   The multi-layered social network

Piotr Bródka, Tomasz Filipowski, Przemysław Kazienko

**Output**: The list of groups within the *MSN*

1. Calculate the *CLECC*($x,y,\alpha$) for each pair ($x,y$) where $x \in MN(y)$ and selected $\alpha$
2. Remove all edges between par ($x,y$) for which the CLECC was the lowest. In case there are two pairs with the lowest CLECC select randomly one of them.
3. Recalculate the *CLECC*($x,y,\alpha$) for all $z:z \in MN(x) \cup MN(y)$ and selected $\alpha$
4. If the deletion of edges will lead to the separation of network into the subgraphs, validate them against selected condition for the group existence (in the original *MSN*). If the subgraph is a multi-layered group do not remove any more edges.
5. Repeat from step 2 until there are only groups or single nodes.

This approach looks similar to approaches presented by Girvan and Newman [3], and Radicchi [7] but by using the new *CLECC measure* it allow us to extract multi-layered communities in multi-layered social networks.

## 6    Conclusions and Future Work

In this paper, the general concept of the new measure called multi-layered edge clustering coefficient was presented. This measure can be used to detect how strongly are connected the neighbourhoods of two nodes linked by an edge. In consequence, this measure can be utilized to detect communities in the multi-layered social network.

Future work will focus on complex examination of the proposed measure and method. Additionally, the new version of *CLECC*, which will also take into account the weights and directions of the *MSN* edges, will be developed.

**Acknowledgments.** The work was supported by: Fellowship co-Financed by the European Union within the European Social Fund, The Polish Ministry of Science and Higher Education, the development project, 2009-2011 and the research project, 2010-13, and the training in the "Green Transfer" project co-financed by the European Union from the European Social Fund